\documentclass[
  twoside,
  twocolumn,
  prl,
  showpacs,
  floatfix,
]{revtex4-1}

\usepackage[dvips]{graphicx}
\usepackage{color}
\usepackage{subfigure}
\usepackage{amsmath}
\usepackage{times}

\newcommand{\eV}{\text{eV}}
\newcommand{\cm}{\text{cm}}

\newcommand{\fig}[1]{Fig.~\ref{#1}}

\newcommand{\eq}[1]{Eq.~(\ref{#1})}
\begin{document}

\title{
  The Urbach tail in silica glass from first principles
}

\author{B. Sadigh, P. Erhart, D. {\AA}berg, A. Trave, E. Schwegler, and J. Bude}
\affiliation{
  Lawrence Livermore National Laboratory,
  Chemistry, Materials and Life Sciences Directorate,
  Livermore, CA, 94550
}

% -    42.70.Ce  --  optical materials: glasses, quartz
% -  - 71.15.Mb  --  DFT, LDA, GGA
%    - 71.23.An  --  Electronic structure of disordered solids: Theories and models; localized states
%      71.23.Cq  --  Electronic structure of disordered solids: Amorphous semiconductors, metallic glasses, glasses 
% -    71.55.Jv  --  Disordered structures; amorphous and glassy solids (localization in disordered solids)
% *  - 78.20.Bh  --  Optical properties of bulk materials and thin films: Theory, models, and numerical simulation 
%    * 78.40.Pg  --  Absorption and reflection spectra: visible and ultraviolet: Disordered solids 
%\pacs{ 78.20.Bh, 71.55.Jv, 42.70.Ce, 71.15.Mb }

\date{\today}

\begin{abstract}
We present density-functional theory calculations of the optical absorption spectra of silica glass for temperatures up to 2400\,K. The calculated spectra exhibit exponential tails near the fundamental absorption edge that follow the Urbach rule, in good agreement with experiments. We also discuss the accuracy of our results by comparing to hybrid exchange correlation functionals. By deriving a simple relationship between the exponential tails of the absorption coefficient and the electronic density-of-states, we establish a direct link between the photoemission and the absorption spectra near the absorption edge. This relationship is subsequently employed to determine the lower bound to the Urbach frequency regime. Most interestingly, in this frequency interval, the optical absorption is Poisson distributed with very large statistical fluctuations. Finally, We determine the upper bound to the Urbach frequency regime by identifying the frequency at which transition to Poisson distribution takes place.
\end{abstract}

\maketitle

At finite temperatures, the absorption spectra of insulators can be modified substantially through interaction of the electronic states with lattice vibrations. In 1953 Urbach \cite{Urb53} observed an exponential energy dependence of the absorption coefficient near the fundamental absorption edge that varied with temperature as follows
\begin{align}
  \overline{\alpha}(\omega,T) = 
  \alpha_0\exp\left[-\sigma\frac{\hbar\omega_0(T)-\hbar\omega}{kT}\right].
  \label{eq:urbach}
\end{align}
Here $\omega_0(T)$ is a linear function of temperature, which at zero Kelvin is defined to be the optical gap, and $\sigma$ and $\alpha_0$ are constants that can be extracted from experiments. The so-called Urbach rule described by \eq{eq:urbach}, has been observed universally in crystals as well as glasses, in both semiconductors and insulators. Its origin has been discussed extensively over the years \cite{Kei66, SouCohEco84, HalLax66} which has lead to the consensus that it arises from transitions between localized electronic levels resulting from temporal fluctuations of the band-edge electrons into the band gap and extended band-like states. Since Urbach behavior involves electron localization \cite{AbtDra07, PanInaZha08} in the presence of vibrations, it is not obvious that standard electronic structure theories such as the density-functional theory (DFT) in the local-density (LDA) or gradient-corrected (GGA) approximations can capture these effects, in particular when considering their systematic underestimation of band gaps of insulators. In a pioneering work, Drabold {\it et al.} performed {\it ab-initio} LDA molecular-dynamics (MD) simulations of amorphous Si and calculated the fluctuations in the single-particle energies at the band edges \cite{DraFedKle91}. These were found to be in good agreement with the band tail widths deduced from photoemission spectra demonstrating the applicability of {\it ab-initio} MD in the adiabatic approximation to describe the electronic structure of semiconductors at finite temperatures. The quantitative computation of the Urbach tail of the optical absorption, however, remains a daunting task. It requires calculating the probability distribution of rare dipole transition events, which necessitates long-time {\it ab-initio} MD simulations in order to obtain the time-dependent fluctuations of single-particle energies and dipole matrix elements. In the past, almost invariably the assumption has been made that the latter do not vary appreciably with atomic displacements or frequency, an assumption that is very difficult to justify especially at high temperatures.

In this paper, we investigate the Urbach rule in defect-free silica glass using {\it ab-initio} MD simulations. This has been motivated by the need to develop a better understanding of the process of laser damage to silica optics, which is of importance to diverse fields ranging from opto-electronics to inertial confinement fusion. In the past, a considerable amount of computational studies has been directed toward understanding zero Kelvin absorption due to defects in silica \cite{PacIer97, BakRasPan04, AlkBroPas08}. Recently, the role of temperature has been emphasized by experiments where damage was generated far below the bulk material threshold by photons of energy 3.55\,eV when silica was heated to about 2200\,K \cite{BudGusMat07}. The Urbach rule plays a crucial role here since the exponential dependence of absorption on temperature in \eq{eq:urbach}, necessitates the existence of a critical temperature $T_c$, at which the glass absorbs more photon energy than it can dissipate leading to thermal run-away and macroscopic damage \cite{BudGusMat07}. However, extrapolation to higher temperatures of the experimental spectra available up to 1900\,K \cite{SaiIku00} predicts a $T_c$ that is several hundred degrees higher than the measured value. Therefore better understanding of the kinetics of absorption at finite temperatures in the Urbach regime is needed. The objective of this Letter is to study the physical processes that lead to absorption in a temperature and energy range for which experiments are not available. The key finding is that in the Urbach regime absorption occurs locally in space as a Poisson process at sub-terahertz frequencies. On the basis of this work, models that enable quantitative predictions of $T_c$ can be obtained.

\begin{figure}
\includegraphics[scale=0.56]{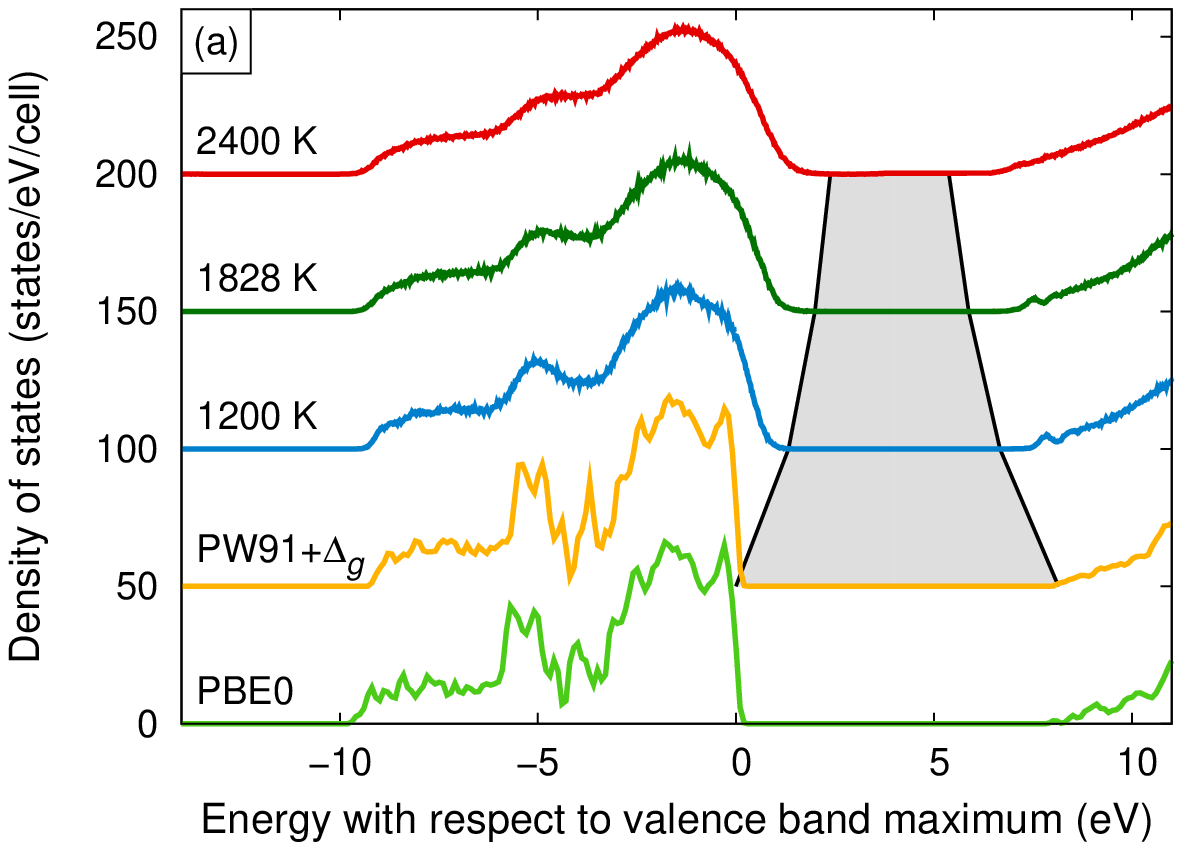}
\includegraphics[scale=0.56]{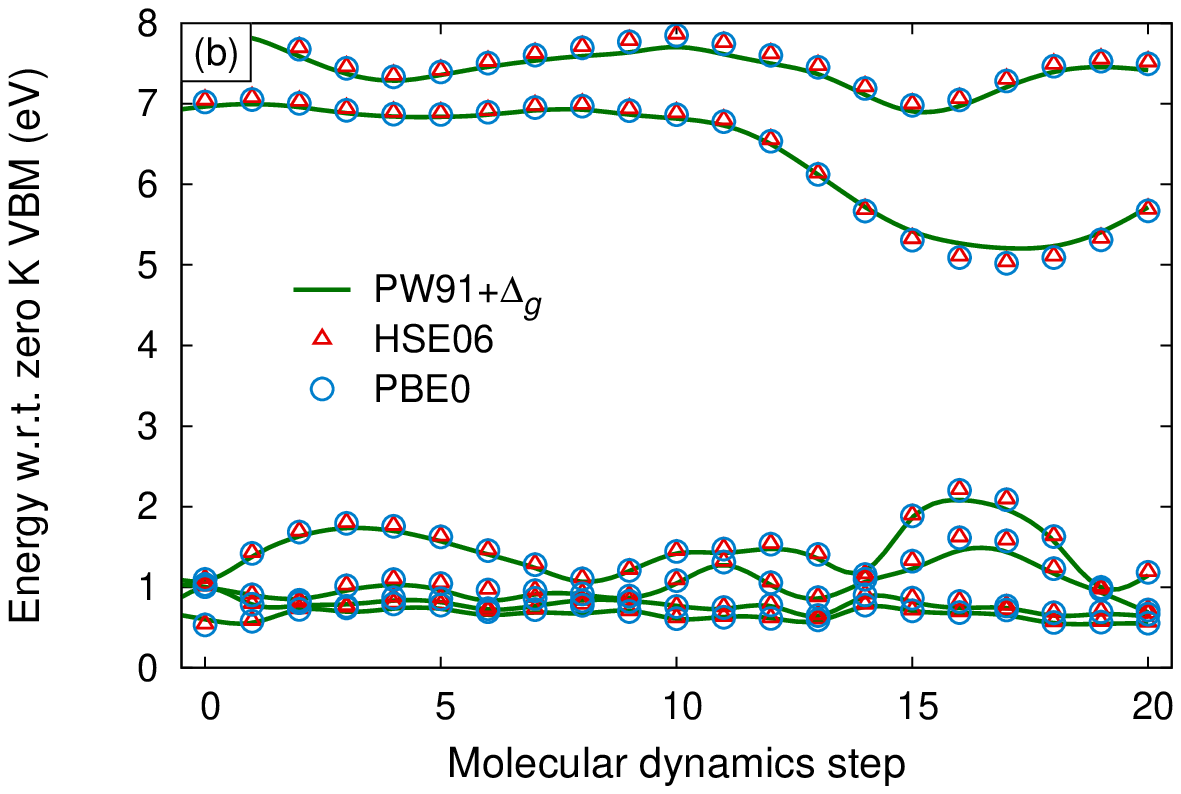}
  \caption{
    (a) Density-of-states for the perfect glass at zero Kelvin calculated using the PBE0 hybrid functional (lowermost line) and the PW91 functional with scissors correction (PW91+$\Delta_g$, second line from the bottom). Also shown is the density of states obtained from MD simulations at three different temperatures using PW91+$\Delta_g$.
    (b) Single-particle energies of the four topmost valence band states and the two lowermost conduction band state as a function of time along an MD trajectory obtained using two different hybrid functionals (HSE06, PBE0) in comparison with the PW91+$\Delta_g$ approach.
  }
  \label{fig:dos}
\end{figure}

The MD simulations presented in this work are performed within the DFT-GGA framework using the PW91 parametrization \cite{PerWan86, Per91} as implemented in the Vienna ab-initio simulation package \cite{VASP} using the projector augmented wave method \cite{PAW}. All calculations involve supercells containing 24 SiO$_2$ formula units and the Brillouin zone is sampled by a $2\times2\times2$ Monkhorst-Pack $k$-point grid. In order to obtain a realistic glass model, we started from a liquid silica model obtained previously \cite{TraTanSca02}, which was quenched down to zero Kelvin over a period of about 10\,ps. The examination of the electronic structure of the resulting configurations revealed defect states due to the presence of stretched and broken bonds. The defect states were eliminated from the model by optimization via a bond-switching Monte Carlo (BSMC) technique \cite{WooWinWea85}. Several BSMC-refined configurations were generated, each representing a random network with fixed bond lengths and angles. Subsequently, the configurations were structurally relaxed to the local GGA total energy minimum. The final glass model that was chosen from this set had preserved its bond lengths and angles after the relaxation process. The electronic density-of-states (DOS) of this configuration is shown in \fig{fig:dos}(a) in comparison with the PBE0 hybrid functional, which incorporates 25\%\ exact exchange \cite{PerBurErn96, PBE0}. The GGA DOS includes a band gap shift of $\Delta_g=2.6\,\eV$ to account for the systematic underestimation of the band gap. The excellent agreement between the two calculations is not surprising since in general it is expected that GGA-DFT, with the exception of the the band gap, can reproduce most of the features of the electronic structure of insulators accurately. However, as mentioned earlier, the electronic states at the band edges undergo localization when they fluctuate inside the band gap. If the magnitude of the GGA-DFT band gap shift depends strongly on the degree of localization of these states, then it would not be possible to develop a quantitative prediction of the Urbach tail with this level of theory. In order to address this issue, we have explicitly compared the time evolution of the band edge states at 2200\,K, calculated from GGA with PBE0 as well as HSE06 \cite{HeyScuErn03} (a hybrid functional that includes screened exchange) calculations, see \fig{fig:dos}(b). It appears that with a constant band gap shift (2.6 \,eV for GGA and 0.9\,eV for HSE06), all calculations can be brought in agreement with each other. We thus conclude that DFT-GGA provides a reasonable basis for modeling the Urbach tail from first principles. A similar conclusion was reached by Alkauskas {\it et al.} in studying point defects \cite{AlkBroPas08}.

The absorption coefficient for photons of energy $\hbar \omega$ of an atomic configuration $\boldsymbol{X}$, can be calculated as follows
\begin{align}
  \label{eq:alpha}
  \alpha(\omega;\boldsymbol{X})
  &= \sqrt{2}\frac{\omega}{c}\sqrt{\left|
    \epsilon(\omega;\boldsymbol{X})\right|-\epsilon_R(\omega;\boldsymbol{X})},
\end{align}
where $\epsilon(\omega;\boldsymbol{X})$ is the complex dielectric function $\epsilon = \epsilon_R + i \epsilon_I$. In the velocity gauge, $\epsilon_I$ can be directly computed from the single-particle wave functions and energies \cite{DelGir93, LevAll91} as well as their occupancies $f_{n\boldsymbol{k}}$ as follows
\begin{align}
  \epsilon_I(\omega;\boldsymbol{X})
  &= \frac{4\pi^2e^2}{m_e^2\omega^2}\sum_{n,n'} 
  (f_{n'\boldsymbol{k}}-f_{n\boldsymbol{k}})\left|
  M^{\boldsymbol{k}}_{nn'}(\boldsymbol{X})\right|^2\nonumber\\ 
  &\quad \times \delta\left(
  \Delta_g+e_{n'\boldsymbol{k}}(\boldsymbol{X})
  -e_{n\boldsymbol{k}}(\boldsymbol{X})-\hbar\omega\right),
  \label{eq:eps}
\end{align}
where $M^{\boldsymbol{k}}_{nn'}(\boldsymbol{X})$ are the polarization-averaged dipole matrix elements between the states $n\boldsymbol{k}$ in the valence band and $n'\boldsymbol{k}$ in the conduction band. The sums in \eq{eq:eps} run over bands and spins, and the real part $\epsilon_R$ can be obtained from $\epsilon_I$ through a Kramers-Kronig relation. Since the latter involves an integration over the entire frequency spectrum, we have included as many as 1000 unoccupied bands in our calculations in order to obtain accurate values for $\epsilon_R$.

\begin{figure}
\includegraphics[scale=0.56]{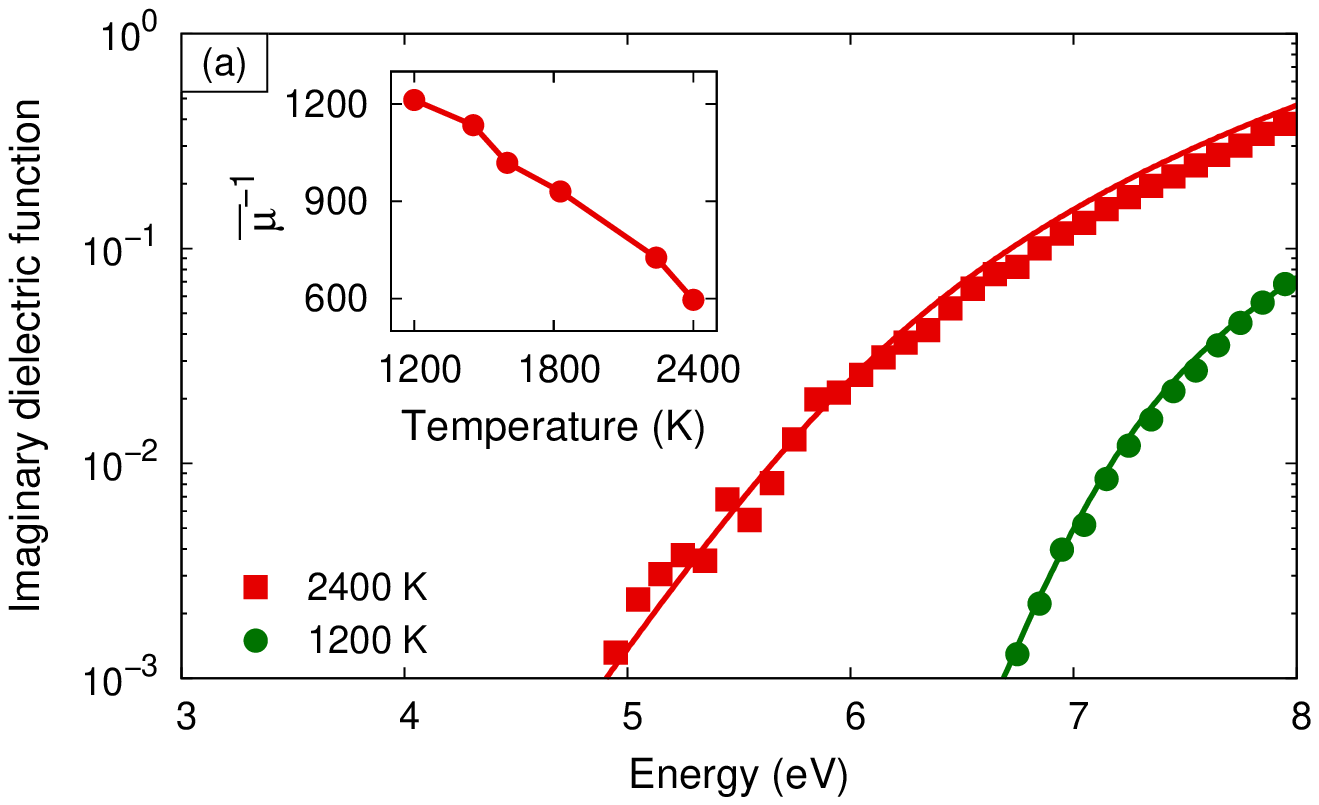}
\includegraphics[scale=0.56]{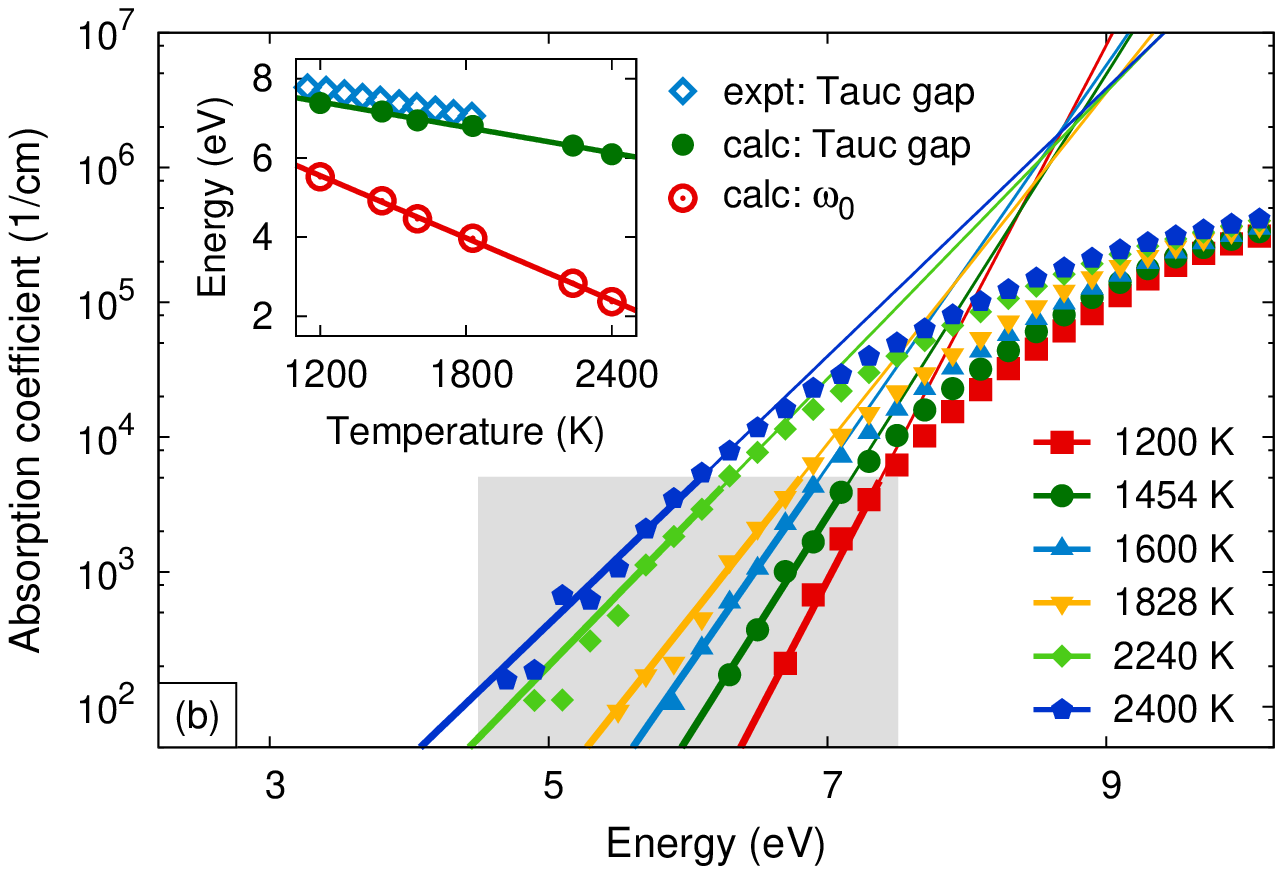}
  \caption{
    (a) JDOS rescaled by $\overline{\mu}(T)$ (solid lines) and imaginary dielectric function (symbols) at two different temperatures calculated from MD simulations. The inset shows $\overline{\mu}(T)^{-1}$ as a function of temperature.
    (b) Absorption coefficient in the Urbach regime at different temperatures calculated from MD simulations using the PW91 functional with a scissors shift. The inset shows the result of the fit to \eq{eq:urbach} and the comparison with the experimental data from Saito and Ikushima \cite{SaiIku00}. The shaded region depicts the Urbach regime. 
  }
  \label{fig:abs}
\end{figure}

At finite temperatures, the response functions as well as the DOS are calculated by  classical ensemble averaging over ionic displacements in the Born-Oppenheimer approximation, which amounts to averaging over the MD simulation time steps. In this way, the electronic transitions are treated as instantaneous. The finite-temperature electronic state occupancies are determined by the Fermi-Dirac distribution $f_{n\boldsymbol{k}} = 1/\left[1+\exp\left(e_{n\boldsymbol{k}}-\mu(T)/k_B T\right)\right]$. The electron chemical potential $\mu(T)$ is calculated from the charge neutrality condition, $\int_{-\infty}^{\mu(T)}\left<\rho(\epsilon)\right>_T~d\epsilon = N_e$, where $N_e$ is the total number of electrons, and $\left<\rho(\epsilon)\right>_T$ is the average DOS. Figure~\ref{fig:dos}(a) shows the average DOS at three different temperatures. The gray region depicts the band gap narrowing with increasing temperature, which contributes to creating holes and electrons in the valence and the conduction bands. The equilibrium concentrations of free electrons as a function of temperature can be calculated by summing up the total occupancies of the conduction band states. Although the free electron concentration can be as large as $10^{17}\,\cm^{-3}$ at 2400\,K, it is still too small to have any measurable impact on the absorption coefficients in the Urbach regime. The latter can thus be calculated using zero Kelvin occupancies and thus neglecting free-electron absorption. The DOS is connected to absorption through the joint density-of-states (JDOS), which for zero Kelvin occupancies is defined as $J(\omega) = \int\rho_v(\omega')\rho_c(\omega'+\omega)~d\omega'$, where $\rho_{v(c)}(\omega)$ is the DOS of the valence (conduction) bands. At finite temperatures, a direct relationship between the JDOS and the DOS only exists if the fluctuations in the valence and the conduction bands are independent
\begin{align}
  \label{eq:jdos}
  \left<\mathcal{J(\omega)}\right>_T \approx 
  \int\left<\rho_v(\omega')\right>_T\left<
  \rho_c(\omega'+\omega)\right>_T~d\omega'.
\end{align}
We find that the above is indeed a very good approximation over the entire frequency range as illustrated in \fig{fig:abs}(a), which shows the structure of the low-energy exponential tail of the JDOS for two temperatures. The figure also illustrates that the low-frequency exponential tails of dielectric function and JDOS coincide when the latter is scaled by a temperature-dependent effective dipole transition probability $\overline{\mu}(T)$,
\begin{align}
  \label{eq:jj}
  \left<\epsilon_I(\omega)\right>_T
  \approx \overline{\mu}(T)\left<\mathcal{J}(\omega)\right>_T.
\end{align}
This important result suggests that exact knowledge of the matrix elements is not necessary to reproduce the frequency dependence of the low-energy exponential tail of the dielectric function. They cannot, however, be neglected entirely, since they lead to an effective temperature-dependent coefficient, the inverse of which decreases linearly with temperature and changes by a factor of two between 1200 and 2400\,K, as shown in the inset of \fig{fig:abs}(a).

After obtaining the dielectric functions at different temperatures, it is straightforward to calculate the absorption spectra according to \eq{eq:alpha}, which leads to the curves in \fig{fig:abs}(b). Fitting the low-energy exponential tails of these spectra to \eq{eq:urbach}, yields a linear temperature dependence for $\omega_0(T)$ in accordance with Urbach's rule, as shown in the inset of \fig{fig:abs}(b). Furthermore, we obtain $\sigma=0.473$ in fair agreement with the experimental value of $\sigma=0.585$ reported by Saito and Ikushima \cite{SaiIku00}. The data also enable us to calculate the Tauc gap, which is the threshold energy for the onset of extended-to-extended electronic transitions. Following Saito and Ikushima \cite{SaiIku00} who defined the Tauc gap as the photon energy corresponding to $5\times10^3\,\cm^{-1}$ absorption, one obtains the data shown in the inset of \fig{fig:abs}(b) demonstrating excellent agreement. The calculated linear temperature dependence of $\omega_0(T)$ in \fig{fig:abs}(b) extends the Urbach rule up to 2400\,K.

It is interesting to note that in the Urbach regime the ratio $\left<\epsilon_I(\omega)\right>_T/\left<\epsilon_R(\omega)\right>_T\ll 1$. A first order Taylor expansion of \eq{eq:alpha} with respect to this quantity yields the following expressions for the absorption coefficient
\begin{align}
  \label{eq:abs}
  \left<\alpha(\omega)\right>_T
  \approx
  \frac{\omega}{c}\frac{\left<\epsilon_I(\omega)\right>_T}
       {\sqrt{\left<\epsilon_R(\omega)\right>_T}}
       \approx
       \frac{\omega}{c}\frac{\overline{\mu}(T)}
	    {\sqrt{\left<\epsilon_R(0)\right>_T}}
	    \left<\mathcal{J}(\omega)\right>_T.
\end{align}
The second approximation above is obtained by a zeroth order expansion about the static dielectric constant $\left<\epsilon_{R}(0)\right>_T$, and utilizes our earlier finding that the Urbach tail of the imaginary dielectric function can be obtained from the JDOS, see Eq.~\eq{eq:jj}. Using Eqs.~\eq{eq:jdos} and \eq{eq:abs}, we can establish a simple relationship between the absorption coefficient and the DOS in the vicinity of the absorption edge, where the temperature dependence of the prefactor mainly enters through the effective dipole-transition probability $\overline{\mu}(T)$, while $\left<\epsilon_{R}(0)\right>_T$ varies only weakly with temperature, i.e. from 1.81 at 0\,K to 1.99 at 2400\,K, corresponding to an increase of less than 10\%. We point out that the DOS is obtained experimentally through photoemission spectroscopy. Therefore, the above result provides a direct link between the photoemission and the optical absorption experiments in the Urbach tail region of the spectrum.

\begin{figure}
\includegraphics[scale=0.56]{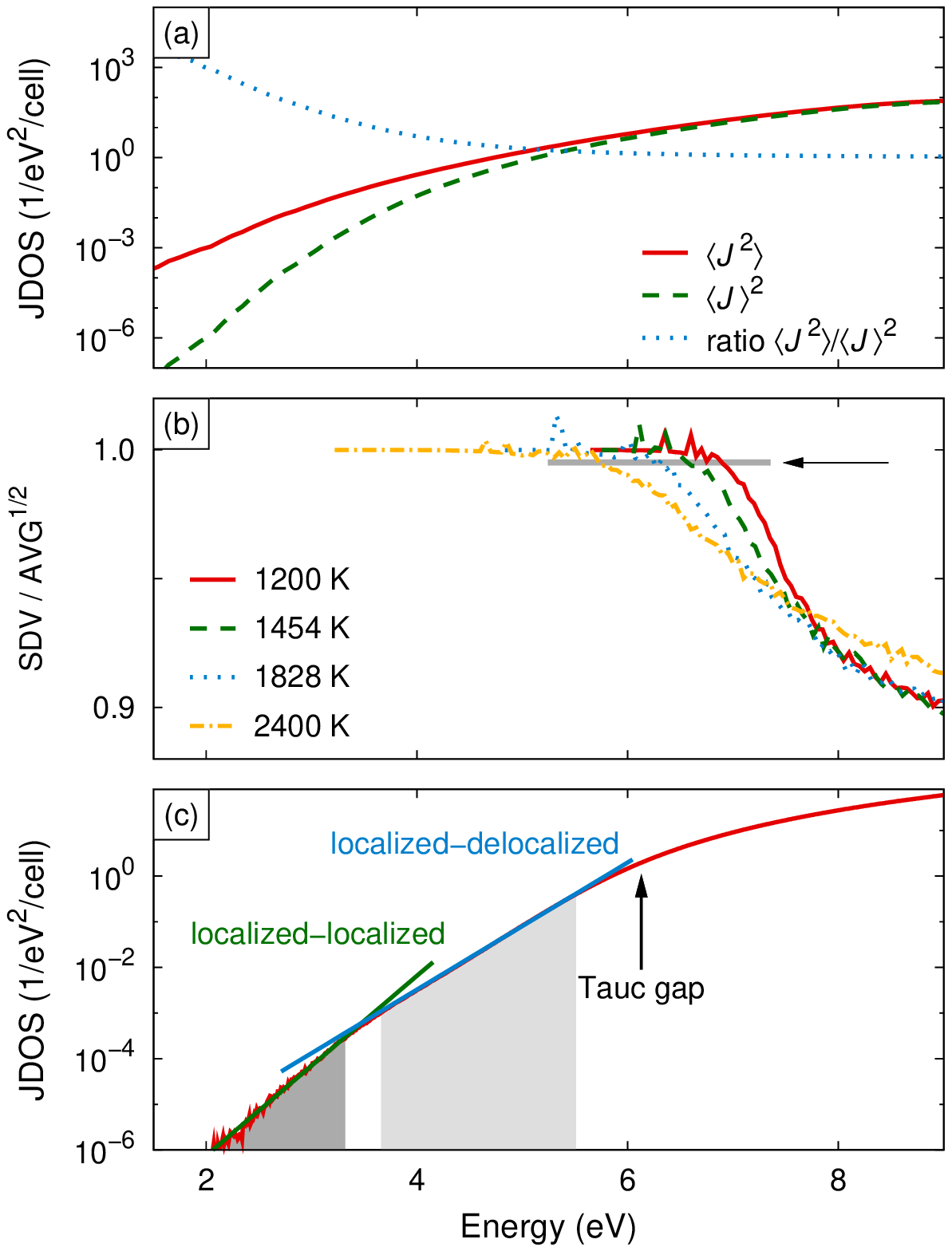}
\includegraphics[scale=0.56]{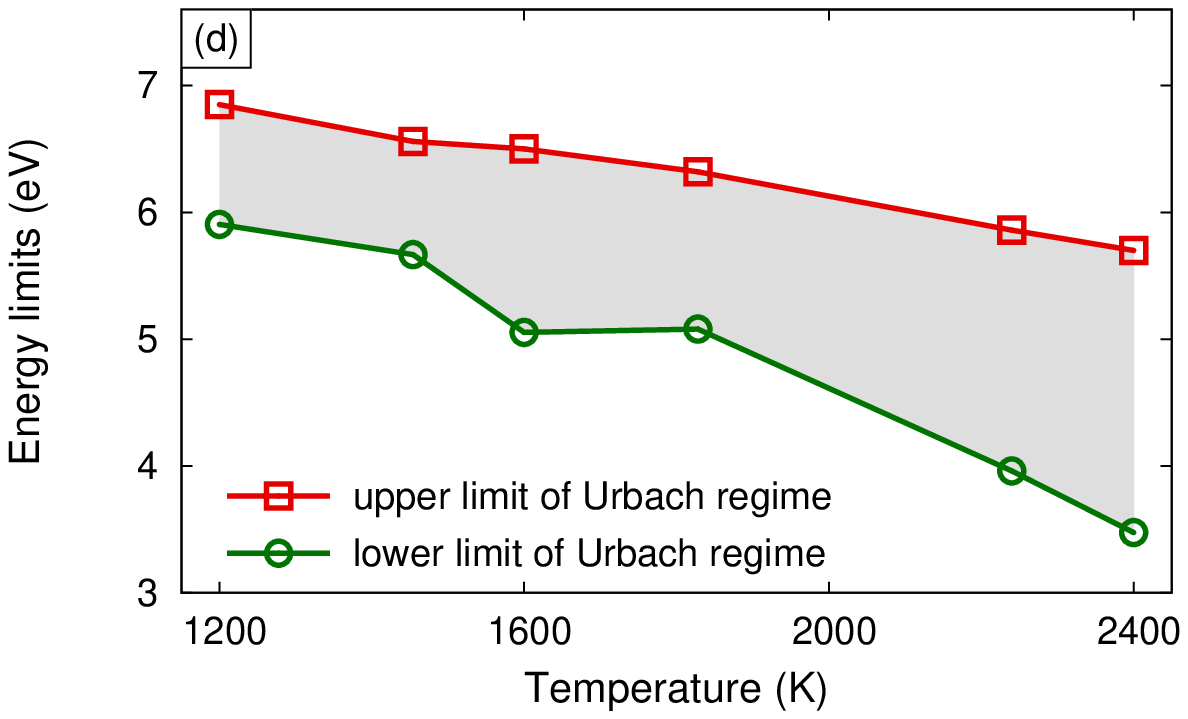}
  \caption{
    (a) First and second moment of the joint density of states (JDOS) as well as their ratio at 2400\,K.
    (b) In the Urbach regime The standard deviation of the JDOS equals the square root of the average indicative of a Poisson distribution. The energy at which the ratio deviates from one provides a simple measure for the upper limit of the Urbach regime as indicated by the arrow.
    (c) At low energies, the JDOS exhibits two distinct regions corresponding to localized-delocalized (Urbach regime) and localized-localized transitions. Each region is described by a different exponential the crossing point of which defines a lower limit for the Urbach regime. Note that the Tauc gap lies above the region within which the JDOS exhibits exponential tails.
    (d) Upper and lower limits of the Urbach regime extracted from the data presented in (b) and (c).
  }
  \label{fig:j2}
\end{figure}

Equation~(\ref{eq:abs}) has the important implication that at finite temperatures the dipole matrix elements average out such that in the Urbach tail they can be replaced by a temperature-dependent prefactor. The atomic vibrations also lead to significant statistical fluctuations, which at present are computationally too expensive for us to quantify accurately. However, we can use \eq{eq:abs} to obtain a lower bound on the statistical fluctuations in $\alpha(\omega)$ by neglecting the fluctuations in the matrix elements and study $\left<\mathcal{J}^2(\omega)\right>_T$. This is shown in \fig{fig:j2}(a), where comparison is made with $\left<\mathcal{J}(\omega)\right>^2_T$. A significant departure between the two curves is observed in the Urbach tail, where the ratio $\left<\mathcal{J}^2(\omega)\right>_T/ \left<\mathcal{J}(\omega)\right>^2_T$ can reach $10^4$. Hence the standard deviation from mean absorption for frequencies in the Urbach tail is up to two orders of magnitude larger than $\left<\alpha(\omega)\right>_T$.

The large fluctuations in the Urbach regime originate from the discrete nature of the JDOS itself. Even if $\left<\mathcal{J}(\omega)\right>_T\ll 1$, there can only exist an integer number $\mathcal{N}$ of pairs of electronic states available for transition at each instant of time. Hence the absorption coefficient locally fluctuates between zero and $\overline{\mu}^2\times \mathcal{N}$, which can amount to fluctuations much larger than the mean value $\left<\mathcal{J}\right>_T$ itself. This behavior is a consequence of the quantum nature of matter. Let us define an absorption event as the instant of time when $\mathcal{N}>0$. Whenever absorption events occur so rarely that they can be considered independent, the absorption process is Poisson distributed, with $\left<\mathcal{J}(\omega)\right>_T$ interpreted as the average rate of occurrence. An important signature of the Poisson distribution is that its standard deviation is equal to the square root of its average. As shown in \fig{fig:j2}(b) in the Urbach regime, this equality indeed holds while the ratio rapidly decreases at only slightly higher energies. The sharp transition provides a natural definition for the frequency $\omega_U$ which signifies the upper bound to the Urbach region and below which the statistics become Poisson distributed. Figure~\ref{fig:j2}(d) shows that in the temperature range considered here, $\omega_U$ decreases linearly. These observations bestow the finite-temperature JDOS in the Urbach tail with distinct physical significance as it represents the average rate of occurrence of absorption events in this frequency interval.

The Urbach tail region is a closed frequency interval in the optical spectrum. Above, we have determined the upper bound for this interval. We can also determine its lower bound using \eq{eq:jdos} to compute $\left<\mathcal{J}(\omega)\right>_T$ down to very small values with good statistical accuracy. Figure~\ref{fig:j2}(c) shows that at very low energies there is a transition in the signature exponential decay of the Urbach tail to a steeper decay curve. The frequency $\omega_L$ at which this transition occurs can be used to identify the lower bound of the Urbach region, the temperature dependence of which is shown in \fig{fig:j2}(d). The optical transitions in this lower-frequency region correspond to transitions between localized levels in the exponential tails of the valence and the conduction bands while the Urbach tail originates from transitions between localized tail states and extended band-like states. 

Electronic localization in the Urbach tail has been discussed extensively in the literature \cite{SouCohEco84, HalLax66, AbtDra07, PanInaZha08}. As will be described in detail in a future publication, we indeed observe pronounced localization in our simulations. Here it suffices to point out that in the Urbach region the conduction band edge is practically flat in reciprocal space which is is a manifestation of the localization of these eigenstates within the 72-atom supercell. In contrast, the conduction band minimum of the perfect glass exhibits significant dispersion in reciprocal space with a standard deviation of 250\,meV. 

Finally, let us discuss the impact of the above findings on the modeling of laser heating in silica. Commonly, this process is described by a heat conduction equation with a source term that incorporates energy deposition by linear coupling to the laser light, $\overline{\alpha}\left(\omega,T\right)I({\bf r})$, where $I({\bf r})$ is the laser light intensity. Neglecting fluctuations, this term can be parametrized by $\overline{\alpha}(\omega,T) = \left<\alpha(\omega)\right>_T$. However, the rare event nature of absorption in the Urbach regime calls for $\overline{\alpha}(\omega,T)$ to be treated as a discrete Poisson process, where at a rate proportional to $\left<\mathcal{J}(\omega)\right>_T$ an absorption event of strength $\omega \overline{\mu}^2/ c \sqrt{\left<\epsilon_R(0)\right>_T}$ takes place. The large fluctuations introduced in this way can reduce the predicted thermal run-away temperature leading to better agreement with experiments.

This work performed under the auspices of the U.S. Department of Energy by Lawrence Livermore National Laboratory under Contract DE-AC52-07NA27344 with support from the Laboratory Directed Research and Development Program.


\begin{thebibliography}{28}
\expandafter\ifx\csname natexlab\endcsname\relax\def\natexlab#1{#1}\fi
\expandafter\ifx\csname bibnamefont\endcsname\relax
  \def\bibnamefont#1{#1}\fi
\expandafter\ifx\csname bibfnamefont\endcsname\relax
  \def\bibfnamefont#1{#1}\fi
\expandafter\ifx\csname citenamefont\endcsname\relax
  \def\citenamefont#1{#1}\fi
\expandafter\ifx\csname url\endcsname\relax
  \def\url#1{\texttt{#1}}\fi
\expandafter\ifx\csname urlprefix\endcsname\relax\def\urlprefix{URL }\fi
\providecommand{\bibinfo}[2]{#2}
\providecommand{\eprint}[2][]{\url{#2}}

\bibitem[{\citenamefont{Urbach}(1953)}]{Urb53}
\bibinfo{author}{\bibfnamefont{F.}~\bibnamefont{Urbach}},
  \bibinfo{journal}{Phys. Rev.} \textbf{\bibinfo{volume}{92}},
  \bibinfo{pages}{1324} (\bibinfo{year}{1953}).

\bibitem[{\citenamefont{Keil}(1966)}]{Kei66}
\bibinfo{author}{\bibfnamefont{T.~H.} \bibnamefont{Keil}},
  \bibinfo{journal}{Phys. Rev.} \textbf{\bibinfo{volume}{144}},
  \bibinfo{pages}{582} (\bibinfo{year}{1966}).

\bibitem[{\citenamefont{Soukoulis et~al.}(1984)\citenamefont{Soukoulis, Cohen,
  and Economou}}]{SouCohEco84}
\bibinfo{author}{\bibfnamefont{C.~M.} \bibnamefont{Soukoulis}},
  \bibinfo{author}{\bibfnamefont{M.~H.} \bibnamefont{Cohen}}, \bibnamefont{and}
  \bibinfo{author}{\bibfnamefont{E.~N.} \bibnamefont{Economou}},
  \bibinfo{journal}{Phys. Rev. Lett.} \textbf{\bibinfo{volume}{53}},
  \bibinfo{pages}{616} (\bibinfo{year}{1984}).

\bibitem[{\citenamefont{Halperin and Lax}(1966)}]{HalLax66}
\bibinfo{author}{\bibfnamefont{B.~I.} \bibnamefont{Halperin}} \bibnamefont{and}
  \bibinfo{author}{\bibfnamefont{M.}~\bibnamefont{Lax}},
  \bibinfo{journal}{Phys. Rev.} \textbf{\bibinfo{volume}{148}},
  \bibinfo{pages}{722} (\bibinfo{year}{1966}).

\bibitem[{\citenamefont{Abtew and Drabold}(2007)}]{AbtDra07}
\bibinfo{author}{\bibfnamefont{T.~A.} \bibnamefont{Abtew}} \bibnamefont{and}
  \bibinfo{author}{\bibfnamefont{D.~A.} \bibnamefont{Drabold}},
  \bibinfo{journal}{Phys. Rev. B} \textbf{\bibinfo{volume}{75}},
  \bibinfo{pages}{045201} (\bibinfo{year}{2007}).

\bibitem[{\citenamefont{Pan et~al.}(2008)\citenamefont{Pan, Inam, Zhang, and
  Drabold}}]{PanInaZha08}
\bibinfo{author}{\bibfnamefont{Y.}~\bibnamefont{Pan}},
  \bibinfo{author}{\bibfnamefont{F.}~\bibnamefont{Inam}},
  \bibinfo{author}{\bibfnamefont{M.}~\bibnamefont{Zhang}}, \bibnamefont{and}
  \bibinfo{author}{\bibfnamefont{D.~A.} \bibnamefont{Drabold}},
  \bibinfo{journal}{Phys. Rev. Lett.} \textbf{\bibinfo{volume}{100}},
  \bibinfo{pages}{206403} (\bibinfo{year}{2008}).

\bibitem[{\citenamefont{Drabold et~al.}(1991)\citenamefont{Drabold, Fedders,
  Klemm, and Sankey}}]{DraFedKle91}
\bibinfo{author}{\bibfnamefont{D.~A.} \bibnamefont{Drabold}},
  \bibinfo{author}{\bibfnamefont{P.~A.} \bibnamefont{Fedders}},
  \bibinfo{author}{\bibfnamefont{S.}~\bibnamefont{Klemm}}, \bibnamefont{and}
  \bibinfo{author}{\bibfnamefont{O.~F.} \bibnamefont{Sankey}},
  \bibinfo{journal}{Phys. Rev. Lett.} \textbf{\bibinfo{volume}{67}},
  \bibinfo{pages}{2179} (\bibinfo{year}{1991}).

\bibitem[{\citenamefont{Pacchioni and Ieran\'o}(1997)}]{PacIer97}
\bibinfo{author}{\bibfnamefont{G.}~\bibnamefont{Pacchioni}} \bibnamefont{and}
  \bibinfo{author}{\bibfnamefont{G.}~\bibnamefont{Ieran\'o}},
  \bibinfo{journal}{Phys. Rev. B} \textbf{\bibinfo{volume}{56}},
  \bibinfo{pages}{7304} (\bibinfo{year}{1997}).

\bibitem[{\citenamefont{Bakos et~al.}(2004)\citenamefont{Bakos, Rashkeev, and
  Pantelides}}]{BakRasPan04}
\bibinfo{author}{\bibfnamefont{T.}~\bibnamefont{Bakos}},
  \bibinfo{author}{\bibfnamefont{S.~N.} \bibnamefont{Rashkeev}},
  \bibnamefont{and} \bibinfo{author}{\bibfnamefont{S.~T.}
  \bibnamefont{Pantelides}}, \bibinfo{journal}{Phys. Rev. B}
  \textbf{\bibinfo{volume}{70}}, \bibinfo{pages}{075203}
  (\bibinfo{year}{2004}).

\bibitem[{\citenamefont{Alkauskas et~al.}(2008)\citenamefont{Alkauskas,
  Broqvist, and Pasquarello}}]{AlkBroPas08}
\bibinfo{author}{\bibfnamefont{A.}~\bibnamefont{Alkauskas}},
  \bibinfo{author}{\bibfnamefont{P.}~\bibnamefont{Broqvist}}, \bibnamefont{and}
  \bibinfo{author}{\bibfnamefont{A.}~\bibnamefont{Pasquarello}},
  \bibinfo{journal}{Phys. Rev. Lett.} \textbf{\bibinfo{volume}{101}},
  \bibinfo{pages}{046405} (\bibinfo{year}{2008}).

\bibitem[{\citenamefont{Bude et~al.}(2007)\citenamefont{Bude, Guss, Matthews,
  and Spaeth}}]{BudGusMat07}
\bibinfo{author}{\bibfnamefont{J.}~\bibnamefont{Bude}},
  \bibinfo{author}{\bibfnamefont{G.}~\bibnamefont{Guss}},
  \bibinfo{author}{\bibfnamefont{M.}~\bibnamefont{Matthews}}, \bibnamefont{and}
  \bibinfo{author}{\bibfnamefont{M.~L.} \bibnamefont{Spaeth}},
  \bibinfo{journal}{SPIE proceedings} \textbf{\bibinfo{volume}{6720}},
  \bibinfo{pages}{672009} (\bibinfo{year}{2007}).

\bibitem[{\citenamefont{Saito and Ikushima}(2000)}]{SaiIku00}
\bibinfo{author}{\bibfnamefont{K.}~\bibnamefont{Saito}} \bibnamefont{and}
  \bibinfo{author}{\bibfnamefont{A.~J.} \bibnamefont{Ikushima}},
  \bibinfo{journal}{Phys. Rev. B} \textbf{\bibinfo{volume}{62}},
  \bibinfo{pages}{8584} (\bibinfo{year}{2000}).

\bibitem[{\citenamefont{Perdew and Wang}(1986)}]{PerWan86}
\bibinfo{author}{\bibfnamefont{J.~P.} \bibnamefont{Perdew}} \bibnamefont{and}
  \bibinfo{author}{\bibfnamefont{Y.}~\bibnamefont{Wang}},
  \bibinfo{journal}{Phys. Rev. B} \textbf{\bibinfo{volume}{33}},
  \bibinfo{pages}{8800} (\bibinfo{year}{1986}).

\bibitem[{\citenamefont{Perdew}(1991)}]{Per91}
\bibinfo{author}{\bibfnamefont{J.~P.} \bibnamefont{Perdew}}, in
  \emph{\bibinfo{booktitle}{Electronic structure of solids}}, edited by
  \bibinfo{editor}{\bibfnamefont{P.}~\bibnamefont{Ziesche}} \bibnamefont{and}
  \bibinfo{editor}{\bibfnamefont{H.}~\bibnamefont{Eschrig}}
  (\bibinfo{publisher}{Akademie Verlag}, \bibinfo{address}{Berlin},
  \bibinfo{year}{1991}), p.~\bibinfo{pages}{11}.

\bibitem[{\citenamefont{Kresse and Hafner}(1993)}]{VASP}
\bibinfo{author}{\bibfnamefont{G.}~\bibnamefont{Kresse}} \bibnamefont{and}
  \bibinfo{author}{\bibfnamefont{J.}~\bibnamefont{Hafner}},
  \bibinfo{journal}{Phys. Rev. B} \textbf{\bibinfo{volume}{47}},
  \bibinfo{pages}{558} (\bibinfo{year}{1993});
\bibinfo{journal}{{\it ibid.}} \textbf{\bibinfo{volume}{49}},
  \bibinfo{pages}{14251} (\bibinfo{year}{1994});
\bibinfo{author}{\bibfnamefont{G.}~\bibnamefont{Kresse}} \bibnamefont{and}
  \bibinfo{author}{\bibfnamefont{J.}~\bibnamefont{Furthm\"uller}},
\bibinfo{journal}{{\it ibid.}} \textbf{\bibinfo{volume}{54}},
  \bibinfo{pages}{11169} (\bibinfo{year}{1996}{\natexlab{a}});
\bibinfo{journal}{Comp. Mater. Sci.} \textbf{\bibinfo{volume}{6}},
  \bibinfo{pages}{15} (\bibinfo{year}{1996}{\natexlab{b}}).

\bibitem[{\citenamefont{Bl\"ochl}(1994)}]{PAW}
\bibinfo{author}{\bibfnamefont{P.~E.} \bibnamefont{Bl\"ochl}},
  \bibinfo{journal}{Phys. Rev. B} \textbf{\bibinfo{volume}{50}},
  \bibinfo{pages}{17953} (\bibinfo{year}{1994});
\bibinfo{author}{\bibfnamefont{G.}~\bibnamefont{Kresse}} \bibnamefont{and}
  \bibinfo{author}{\bibfnamefont{D.}~\bibnamefont{Joubert}},
  \bibinfo{journal}{{\it ibid.}} \textbf{\bibinfo{volume}{59}},
  \bibinfo{pages}{1758 } (\bibinfo{year}{1999}).

\bibitem[{\citenamefont{Trave et~al.}(2002)\citenamefont{Trave, Tangney,
  Scandolo, Pasquarello, , and Car}}]{TraTanSca02}
\bibinfo{author}{\bibfnamefont{A.}~\bibnamefont{Trave}},
  \bibinfo{author}{\bibfnamefont{P.}~\bibnamefont{Tangney}},
  \bibinfo{author}{\bibfnamefont{S.}~\bibnamefont{Scandolo}},
  \bibinfo{author}{\bibfnamefont{A.}~\bibnamefont{Pasquarello}}, ,
  \bibnamefont{and} \bibinfo{author}{\bibfnamefont{R.}~\bibnamefont{Car}},
  \bibinfo{journal}{Phys. Rev. Lett.} \textbf{\bibinfo{volume}{89}},
  \bibinfo{pages}{245504} (\bibinfo{year}{2002}).

\bibitem[{\citenamefont{Wooten et~al.}(1985)\citenamefont{Wooten, Winer, and
  Weaire}}]{WooWinWea85}
\bibinfo{author}{\bibfnamefont{F.}~\bibnamefont{Wooten}},
  \bibinfo{author}{\bibfnamefont{K.}~\bibnamefont{Winer}}, \bibnamefont{and}
  \bibinfo{author}{\bibfnamefont{D.}~\bibnamefont{Weaire}},
  \bibinfo{journal}{Phys. Rev. Lett.} \textbf{\bibinfo{volume}{54}},
  \bibinfo{pages}{1392} (\bibinfo{year}{1985}).

\bibitem[{\citenamefont{Perdew et~al.}(1996)\citenamefont{Perdew, Burke, and
  Ernzerhof}}]{PerBurErn96}
\bibinfo{author}{\bibfnamefont{J.~P.} \bibnamefont{Perdew}},
  \bibinfo{author}{\bibfnamefont{K.}~\bibnamefont{Burke}}, \bibnamefont{and}
  \bibinfo{author}{\bibfnamefont{M.}~\bibnamefont{Ernzerhof}},
  \bibinfo{journal}{Phys. Rev. Lett.} \textbf{\bibinfo{volume}{77}},
  \bibinfo{pages}{3865} (\bibinfo{year}{1996}), \bibinfo{note}{erratum, {\it
  ibid.} \textbf{78}, 1396(E) (1997)}.

\bibitem[{\citenamefont{Ernzerhof et~al.}(1997)}]{PBE0}
\bibinfo{author}{\bibfnamefont{M.}~\bibnamefont{Ernzerhof}},
  \bibinfo{author}{\bibfnamefont{J.~P.} \bibnamefont{Perdew}},
  \bibnamefont{and} \bibinfo{author}{\bibfnamefont{K.}~\bibnamefont{Burke}},
  \bibinfo{journal}{Int. J. Quantum Chem.} \textbf{\bibinfo{volume}{64}},
  \bibinfo{pages}{285} (\bibinfo{year}{1997}).
\bibinfo{author}{\bibfnamefont{M.}~\bibnamefont{Ernzerhof}} \bibnamefont{and}
  \bibinfo{author}{\bibfnamefont{G.~E.} \bibnamefont{Scuseria}},
  \bibinfo{journal}{J. Chem. Phys.} \textbf{\bibinfo{volume}{110}},
  \bibinfo{pages}{5029} (\bibinfo{year}{1999}).
\bibinfo{author}{\bibfnamefont{C.}~\bibnamefont{Adamo}} \bibnamefont{and}
  \bibinfo{author}{\bibfnamefont{V.}~\bibnamefont{Barone}},
  \bibinfo{journal}{{\it ibid.}} \textbf{\bibinfo{volume}{116}},
  \bibinfo{pages}{6158} (\bibinfo{year}{1999}).

\bibitem[{\citenamefont{Heyd et~al.}(2003)\citenamefont{Heyd, Scuseria, and
  Ernzerhof}}]{HeyScuErn03}
\bibinfo{author}{\bibfnamefont{J.}~\bibnamefont{Heyd}},
  \bibinfo{author}{\bibfnamefont{G.~E.} \bibnamefont{Scuseria}},
  \bibnamefont{and}
  \bibinfo{author}{\bibfnamefont{M.}~\bibnamefont{Ernzerhof}},
  \bibinfo{journal}{J. Chem. Phys.} \textbf{\bibinfo{volume}{118}},
  \bibinfo{pages}{8207} (\bibinfo{year}{2003}), \bibinfo{note}{erratum: {\it
  ibid.} {\bf 124}, 219906 (2006)}.

\bibitem[{\citenamefont{Del~Sole and Girlanda}(1993)}]{DelGir93}
\bibinfo{author}{\bibfnamefont{R.}~\bibnamefont{Del~Sole}} \bibnamefont{and}
  \bibinfo{author}{\bibfnamefont{R.}~\bibnamefont{Girlanda}},
  \bibinfo{journal}{Phys. Rev. B} \textbf{\bibinfo{volume}{48}},
  \bibinfo{pages}{11789} (\bibinfo{year}{1993}).

\bibitem[{\citenamefont{Levine and Allan}(1991)}]{LevAll91}
\bibinfo{author}{\bibfnamefont{Z.~H.} \bibnamefont{Levine}} \bibnamefont{and}
  \bibinfo{author}{\bibfnamefont{D.~C.} \bibnamefont{Allan}},
  \bibinfo{journal}{Phys. Rev. B} \textbf{\bibinfo{volume}{43}},
  \bibinfo{pages}{4187} (\bibinfo{year}{1991}).

\end{thebibliography}
\end{document}